\documentclass[12pt]{article}

\usepackage{epsfig}

\newcommand{\be}{\begin{equation}}
\newcommand{\ee}{\end{equation}}
\newcommand{\bea}{\begin{eqnarray}}
\newcommand{\eea}{\end{eqnarray}}
\newcommand{\ba}{\begin{array}}
\newcommand{\ea}{\end{array}}

\def\bbox{{\,\lower0.9pt\vbox{\hrule \hbox{\vrule height 0.2 cm
\hskip 0.2 cm \vrule height 0.2 cm}\hrule}\,}}
\newcommand{\dsl}{\pa \kern-0.5em /}

\font\mybb=msbm10 at 12pt
\def\bb#1{\hbox{\mybb#1}}
\def\bZ {\bb{Z}}
\def\bR {\bb{R}}

\def\appendix#1{
  \addtocounter{section}{1}
  \setcounter{equation}{0}
  \renewcommand{\thesection}{\Alph{section}}
  \section*{Appendix \thesection\protect\indent \parbox[t]{11.15cm}
  {#1} }
  \addcontentsline{toc}{section}{Appendix \thesection\ \ \ #1}
  }

\begin{document}

\begin{flushright}
\small
UG-05-03\\
KCL-MTH-05-02\\
UB-ECM-PF-05/06\\
DAMTP-2005-23\\
{\bf hep-th/0504011}\\
\date \\
\normalsize
\end{flushright}

\begin{center}


\vspace{.7cm}

\centerline{\LARGE {\bf Cosmological D-instantons}} ~ \\
\centerline{\LARGE{\bf and Cyclic Universes}}
\bigskip

\vspace{1.2cm}

{\large E.A.~Bergshoeff${}^*$, A.~Collinucci${}^*$,
D.~Roest${}^\dagger$,}

{\large J.G.~Russo${}^\ddagger$
and P.K.~Townsend${}^{\diamond}$
} \vskip 1truecm

\small
${}^*$\,{Centre for Theoretical Physics, University of
Groningen,\\
   Nijenborgh 4, 9747 AG Groningen, The Netherlands}\\
\vskip .3truecm

${}^\dagger$\,{Department of Mathematics, King's College,\\
London, Strand, London WC2R 2LS}
\vskip .3truecm

${}^\ddagger$\,{Instituci\'o Catalana de Recerca i
Estudis Avan\c{c}ats,\\
Departament ECM, Facultat de F\'isica,\\
Universitat de Barcelona, Spain}
\vskip .3truecm

${}^\diamond$\,{Department of Applied Mathematics and
Theoretical Physics\\
Centre for Mathematical Sciences, University of Cambridge\\
Wilberforce Road, Cambridge, CB3 0WA, UK}
\vspace{.7cm}


{\bf Abstract}

\end{center}

\begin{quotation}

\small

For models of gravity coupled to hyperbolic sigma models, such as the
metric-scalar sector of IIB supergravity,  we show how smooth trajectories
in the `augmented target space' connect FLRW cosmologies to non-extremal
D-instantons through a cosmological singularity. In particular, we find closed
cyclic universes that undergo  an endless sequence of big-bang to big-crunch
cycles separated by instanton `phases'. We also find `big-bounce' universes in
which a collapsing closed universe bounces off its cosmological singularity to become
an open expanding universe.

\end{quotation}

\newpage

\section{Introduction}
\setcounter{equation}{0}

Homogeneous and isotropic cosmological solutions of
gravity coupled to
$N$ scalar fields can be viewed as trajectories in a
Lorentzian-signature `augmented target space' of dimension
$N+1$. If the scalar field target space is the hyperbolic space
$H_N$, it can happen that the augmented target
space is a Milne
universe, and that cosmological singularities
correspond to points
on trajectories at which the Milne horizon is crossed
\cite{Russo:2004am}. In the models considered
in \cite{Russo:2004am}, cosmological
trajectories were geodesics, and hence
straight lines in a Milne patch of
Minkowski `spacetime' \footnote{See \cite{Waldron:2004gg}
for a related observation
in the context of flat $k=0$ cosmologies of
3-dimensional gravity. It was shown
in \cite{Townsend:2004zp}
that cosmological trajectories are {\it always}
geodesics in an appropriate metric
on the augmented target space but it is only in
rather special cases that this
metric is the Milne metric on Minkowski spacetime.}.
In such models the Milne
horizon is typically crossed twice, corresponding
{\it either} to
expansion from a big-bang singularity followed by
contraction to a big-crunch singularity, {\it or} to a big-bang/big-crunch
transition through a
region behind the Milne horizon in which the scale
factor is imaginary. As the
trajectory through the  horizon is smooth in
Minkowski coordinates,
this construction strongly suggests an interpretation in which a
collapsing universe tunnels through a `forbidden' region in field
space to emerge as a re-expanding universe \cite{Russo:2004am}.

The main aim of this paper is to provide further support
for this idea by showing, in a particular class of models,
that the trajectory behind the
Milne horizon corresponds to a solution of the field
equations of the
{\it Euclidean} action; in other words, an instanton.
Of course, the
general idea that a big-bang singularity might be resolved by a
transition to a Euclidean signature solution, or gravitational
instanton, is not new.  Here, however, we are not
actually resolving the singularity in the spacetime metric; we are instead
re-interpreting it as a mere coordinate singularity in a larger
space in which the scalar fields are on the same footing as the
metric. The instanton solutions that we need for this
re-interpretation are also different; they are the `super-extremal'
D-instantons  \cite{Bergshoeff:2004fq} that generalize the (extremal) D-instanton
of Euclidean IIB supergravity  \cite{Gibbons:1995vg}, hence our title.
An amusing by-product of our analysis is that for $k=1$ cosmologies in a certain
subclass of models,
these instanton-cosmology  transitions may link up to yield a
{\it cyclic} universe; i.e. one that expands from a big-bang to a big-crunch,
passes through an instanton `phase' to re-emerge as an
expanding universe that again recollapses to a big-crunch, followed by a
further instanton `phase', {\it ad infinitum}. Another subclass of models,
which include IIB supergravity,  allow `big-bounce' universes that simply
bounce smoothly off the big-crunch singularity without the need
for an instanton `phase'.

We shall simplify our task by taking $N=2$ but we
consider an arbitrary, but {\it finite}, radius of
the scalar field target space.
We shall also keep arbitrary the spacetime dimension $d$. The model
to be considered here has zero scalar potential, so our starting point
(in the conventions of \cite{Bergshoeff:2004fq})
is the Lagrangian density
\be\label{laginitial}
{\cal L} = \sqrt{\epsilon\det g}\left[ R - {1\over2} (\partial \phi)^2 +
  {1\over2}\epsilon \, e^{b\phi}(\partial\chi)^2 \right] \,,
\ee
for $d$-metric $g$ and scalar fields ($\phi,\chi$).
For $\epsilon=-1$,
the scalar fields parametrize a hyperbolic space
$H_2$ of radius $2/b$. For
$d=10$ and $b=2$ we then have a consistent
truncation of IIB supergravity; note that in this context,
$\chi$ is a pseudo-scalar  so that a
solution with non-zero $\chi$ breaks parity. For
the same values of $d$ and $b$ the Lagrangian with $\epsilon=1$ is
the  consistent truncation of the Euclidean IIB
supergravity shown in
\cite{Gibbons:1995vg,Bergshoeff:2004fq}  to
admit D-instanton solutions.
Note that the scalar fields in this Euclidean
action  parametrize the Lorentzian signature
space $adS_2$, so the Euclidean action  is not positive
definite, but can be made positive definite  \cite{Green:1997tv}
by dualization of $\chi$ (followed by the usual  ``conformal rotation''
of Euclidean quantum gravity  \cite{EQG}).

The above Lagrangian density is thus a natural
generalization to arbitrary
spacetime dimension and arbitrary target space radius of the
metric-scalar sector of IIB supergravity, or  its Euclidean
counterpart. As we shall soon see, cosmological
trajectories of this model
correspond  to motion in a Milne universe.
It was observed  in \cite{Russo:2004am} that there is a special
value of the target space radius (and hence of $b$)
for which the motion is geodesic. However, as we show here,
exact cosmological and instanton solutions  can be found for any
target space radius.

\section{Cosmologies and Instantons}
\label{sec.exact}
\setcounter{equation}{0}

To investigate cosmological solutions of our model,
or to find instanton solutions
of its Euclidean version, we make the ansatz
\be\label{ansatz}
ds^2 = \epsilon \left(e^{\alpha\varphi}f\right)^2 d\lambda^2 +
e^{2\alpha\varphi/(d-1)}d\Sigma_k^2 \, ,\qquad
\phi=\phi(\lambda)\, ,\qquad \chi= \chi(\lambda)\, ,
\ee
where $f$ is an arbitrary function of $\lambda$, and
\be
\alpha = \sqrt{d-1\over 2(d-2)}\, .
\ee
The $(d-1)$-metric $d\Sigma_k^2$ is (at least locally) a maximally
symmetric space of positive ($k=1$), negative ($k=-1$) or
zero ($k=0$) curvature. One can choose coordinates
such that
\be
  d\Sigma_k^2 = (1-kr^2)^{-1}dr^2 + r^2 d\Omega_{d-2}^2 \,,
\label{euss}
\ee
where $d\Omega_{d-2}^2$ is an
$SO(d-1)$-invariant metric on the
unit $(d-2)$-sphere.
This ansatz constitutes a consistent
reduction of the original degrees of freedom
to a three-dimensional
subspace, the `augmented target space', with coordinates
$(\varphi,\phi,\chi)$. The full equations of
motion reduce to a set of
equations that can themselves be derived by
variation of the time-reparametrization invariant effective
action
\be
I= {1\over2}\int d\lambda\,\left\{
f^{-1}\left(\epsilon\dot\varphi^2
- \epsilon\dot\phi^2 + e^{b\phi}\dot\chi^2\right) +
2k(d-1)(d-2)fe^{\varphi/\alpha}\right\}\,,
\ee
where the overdot indicates differentiation
with respect to $\lambda$.
For $\epsilon=-1$ we can interpret  $\lambda$
as a time coordinate related to the
time $t$ of FLRW cosmology in standard coordinates by
\be\label{FLRWtime}
dt \propto e^{\alpha\varphi} f d\lambda\, .
\ee
For $\epsilon =1$ the metric has Euclidean
signature and we can interpret
$\lambda$ as imaginary time.

Before proceeding, it is convenient to define new scalar field
variables $(\psi,\theta)$ by
\bea
e^{(b/2)\phi} &=& e^\psi \cos^2(\theta/2) -
\epsilon e^{-\psi}\sin^2(\theta/2)\, , \nonumber \\
e^{(b/2)\phi} \chi &=& b^{-1}\left(e^\psi
+ \epsilon e^{-\psi}\right)\sin\theta\, ,
\eea
to get the new effective action
\bea
I &=& {1\over2} \int\! d\lambda \Bigg\{ {4\over b^2} f^{-1}
\left[ {b^2\over4}\epsilon\dot\varphi^2 -
\epsilon\dot\psi^2  + {1\over4} \left(e^\psi +
\epsilon e^{-\psi}\right)^2 \dot\theta^2\right] + \nonumber\\
&&\ \ \ \ \ \ \ \ \ \ \ \ + \  2k(d-1)(d-2)f e^{\varphi/\alpha}\Bigg\}\, .
\eea
This is just a reparametrization of the target space but it has the
advantage that the new coordinates are globally valid.
Introducing the new scale-factor variable $\eta$ by
\be\label{etavarphi}
\eta^\gamma = 2\gamma (d-1)\, e^{\varphi/(2\alpha)}\, ,
\ee
where
\be
\gamma = 1/(b\alpha)\, ,
\ee
we arrive at the action
\be\label{effL1}
I= {1\over2}\int d\lambda \left\{{4\over b^2}f^{-1}
\left[ \epsilon(\dot\eta/\eta)^2
-\epsilon \dot\psi^2 + {1\over4} \left(e^\psi +
\epsilon e^{-\psi}\right)^2 \dot\theta^2\right]  +
{b^2\over 4} k f \, \eta^{2\gamma} \right\}\, .
\ee
We remark, for future reference,  that the ansatz (\ref{ansatz}) leads
to $\gamma=2/3$ for  $d=10$ IIB supergravity.

Because of the time-reparametrization invariance, we are free to choose the function $f$; each choice of $f$ corresponds to some choice of time parameter. There are two choices that are particularly convenient, and we now consider them in turn.

\subsection{The `Liouville' gauge}

The simplest way to proceed for general $b$ is to make the
gauge choice
\be
f= 4/b^2 \, .
\ee
{}From (\ref{effL1}) one sees that the effective
Lagrangian in this gauge is
\be \label{H2-action}
L = {1\over2}\left[ -\epsilon \dot\psi^2 + {1\over4}
\left(e^\psi + \epsilon e^{-\psi}\right)^2 \dot\theta^2\right]
+ {1\over2}\left[ \epsilon(\dot\eta/\eta)^2 +
k  \eta^{2\gamma}\right] \, .
\ee
Apart from the constraint, the dynamics of the motion
on the target space, which is manifestly geodesic,
is now separated from the dynamics of the scale factor,
which is determined by a equation of Liouville-type; for this reason
we will call this choice of gauge the ``Liouville gauge''.

As $Sl(2;\bR)$ is
the isometry group of both $H_2$ (the target space
of the Lorentzian action) and $adS_2$ (the target
space of the Euclidean action),  there is a conserved
$Sl(2;\bR)$ `angular momentum' $\ell^\mu$, and the
geodesics are such that
\be\label{geod}
\dot\psi^2  - \epsilon {1\over4}
\left(e^\psi + \epsilon e^{-\psi}\right)^2 \dot\theta^2 = \ell^2\, .
\ee
The constraint ($f$ equation of motion) is
\be\label{con1}
(\dot\eta/\eta)^2 = \ell^2 + k\epsilon\,  \eta^{2\gamma}\, .
\ee
We now present the solutions of the equations of motion of (\ref{H2-action}) subject to
the constraint (\ref{geod}) and (\ref{con1}), first for the target space fields and then for the scale factor.

\subsubsection{Target space geodesics}
\label{subsec.geo}

Geodesics on the $H_2$ ($\epsilon=-1$) or $adS_2$ ($\epsilon=1$)
target space are solutions of the field equations of (\ref{H2-action}) for $\psi$ and $\theta$ subject to (\ref{geod}) and
can be classified as follows, according to whether $\ell^2$ is
positive, negative or zero:

\begin{itemize}
\item $\ell^2>0$. For $\epsilon=1$ the solution is
\bea
\sinh\psi &=& \pm \sqrt{1+{q_-^2\over\ell^2}}\,
\sinh\left[\ell\,\left(\lambda -\lambda_0\right)\right]\nonumber\\
\tan\left(\theta - \theta_0\right) &=& \pm {q_- \over\ell}\,
\tanh\left[\ell\,\left(\lambda -\lambda_0\right)\right]\,,
\label{dosa} \eea for constants $\lambda_0$, $\theta_0$ and $q_-$
(this being the integration constant for the super-extremal
D-instanton of \cite{Bergshoeff:2004fq}). For $\epsilon=-1$ the
solution is \bea \cosh\psi &=& \sqrt{1+\frac{q_-^2}{\ell^2}}\,
\cosh\big[\ell\,(\lambda - \lambda_0)\big]\nonumber \\
\tan(\theta-\theta_0) &=& \pm {q_- \over \ell} \, \coth
\left[\ell\left(\lambda-\lambda_0\right)\right]\, . \label{dosb} \eea In the
special case that $q_-=0$ these solutions simplify, for either
choice of the sign $\epsilon$, to \be\label{special} \psi = \pm
\ell(\lambda-\lambda_0)\, , \qquad \theta= \theta_0 \ , \qquad
(\epsilon=\pm1). \ee

\item $\ell^2<0$. In this case only $\epsilon=1$ is possible,
and the solution is
\bea
\sinh\psi &=& \pm \sqrt{{q_{-}^2\over \left(-\ell^2\right)}-1}\,
\sin\left[\sqrt{-\ell^2}\,
\left(\lambda-\lambda_0\right)\right]\nonumber\\
\tan(\theta-\theta_0) &=& \pm
{q_{-}\over \sqrt{-\ell^2}}\,
\tan\left[\sqrt{-\ell^2}\,
\left(\lambda-\lambda_0\right)\right]\,.
\eea

\item $\ell^2=0$. The only solution for $\epsilon=-1$ in
this case is the trivial one for which both $\psi$ and
$\theta$ are constant. For $\epsilon=1$ the solution is
\be
\sinh\psi = \pm q_-\, \left(\lambda-\lambda_0\right)\, ,\qquad
\tan(\theta-\theta_0) = \pm q_-\,
\left(\lambda-\lambda_0\right)\, .
\ee

\end{itemize}

It should be noted that, in each case, the $\pm$ signs for $\psi$
and $\theta$ can be
chosen independently.

\subsubsection{The scale factor}
\label{subsec.scale}

We next turn to the constraint (\ref{con1}). Given $\ell^2$,
this determines $\eta$ as follows

\begin{itemize}
\item $\ell^2 > 0$.
\bea
\eta^{2\gamma} &=& \eta_0^{2\gamma}
\exp{(\pm 2\ell \gamma \lambda)} \,,  \qquad (k=0) \,, \label{one}\\
\eta^{2\gamma} &=& {\ell^2\over
\sinh^2(\ell\gamma \lambda)} \,, \qquad \qquad (k\epsilon =1) \,, \label{two}\\
\eta^{2\gamma} &=& {\ell^2\over
\cosh^2(\ell\gamma \lambda)} \,, \qquad \qquad (k\epsilon=-1) \,, \label{three}
\eea
for some constant $\eta_0$. Note that all $k=\pm 1$ trajectories are asymptotic
to some $k=0$ trajectory near $\eta=0$, as expected since the $\sigma
$-model matter
satisfies the strong energy condition.

\item $\ell^2<0$. In this case there is a solution
only for $k=\epsilon=1$:
\be\label{timelike}
\eta^{2\gamma} = {-\ell^2\over \sin^2\left(\gamma\sqrt{-\ell^2}\,
\lambda\right)} \,, \qquad (k=\epsilon=1) \,,
\ee

\item $\ell^2=0$. In this case there is a solution only
for $k\epsilon\ge0$. If $k=0$ then $\eta$ is constant. Otherwise
\be\label{ellzero}
\eta^{2\gamma} = 1/ (\gamma\lambda)^2 \,, \qquad (k\epsilon=1)\,.
\ee
\end{itemize}

For $\epsilon=k=1$ these solutions yield the
super-extremal ($\ell^2>0$), sub-extremal ($\ell^2<0$) and
extremal ($\ell^2=0$) D-instantons of
\cite{Bergshoeff:2004fq}. For $\epsilon=-1$ they
yield FLRW cosmologies; from (\ref{FLRWtime}) we
see that the standard FLRW time $t$ is related to
the parameter $\lambda$ by
\be
dt \propto \eta^{2\gamma\alpha^2}\, d\lambda\, .
\ee
Given one of above solutions for $\eta^{2\gamma}$
as a function of $\lambda$ we can determine $\lambda$
as a function of $t$ and hence $\eta$ as a function
of $t$. Of most interest here is the behaviour near
$\eta=0$. For example, for $\ell^2>0$ we have
\be
\eta \sim \eta_0\, e^{-\ell \lambda } \,,
\ee
for $\lambda\rightarrow\infty$, as  $\eta\rightarrow 0$.
This yields (for a choice of integration constant such
that $t\rightarrow0$ as $\lambda\rightarrow\infty$)
\be
-t  \propto e^{-2\gamma\alpha^2\ell \lambda}\, .
\ee
Given that we start with a cosmological solution for
negative $t$, this shows that a big-crunch singularity
will be approached as $t\rightarrow 0$. By considering
the behaviour as $\lambda\rightarrow -\infty$ we may
similarly deduce that a cosmological solution for
positive $t$ must have had a big-bang singularity
at $t=0$. In other words, cosmologies with $\ell^2>0$
are incomplete in the sense that they have a beginning
or an end (or both) at finite $t$. We shall see in the following
section how they can be completed.

\subsection{The `Milne' gauge}

Returning to (\ref{effL1}), we make the new gauge choice
\be\label{Milnegauge}
f= {4\over b^2\, \eta^2}\, .
\ee
As the possible choices of $f$ are related by
a redefinition of the independent variable,
we will need to distinguish the independent
variable in this gauge from the
parameter $\lambda$ used previously. Let us call
the new independent variable $\tau$; it is related
to $\lambda$ through the differential equation
\be\label{lamtau}
d\tau = \eta^2(\lambda) \, d\lambda \, ,
\ee
which can be solved for $\tau(\lambda)$ given any of the scale factor solutions
$\eta(\lambda)$ presented above.

In the gauge (\ref{Milnegauge})  the action is
\be
I= \int\! d\tau \, L_\tau \ ,
\ee
where\footnote{Note that $L_\tau d\tau=L d\lambda $, where $L$ is the
  lagrangian in the gauge used previously.}
\be\label{milneL}
L_\tau =
{1\over 2}
 \epsilon \, \left({d\eta\over d\tau}\right) ^2
+{1\over 2}\eta^2 \left[ -\epsilon \left({d\psi\over d\tau}\right) ^2 + {1\over 4} \left(e^\psi +
\epsilon e^{-\psi}\right)^2 \left({d\theta\over d\tau}\right)^2\right]   +
{k\over 2}   \, \eta^{2\gamma-2} \, .
\ee
We observe that for $\epsilon=-1$ the kinetic
term is that of a particle in a 3-dimensional Milne
`universe', so we call this choice of gauge the `Milne' gauge.

The Milne universe is actually just Minkowski space in
an analog of spherical polar coordinates. The cartesian
field variables $X_\mu$ ($\mu=0,1,2$) are
\bea\label{cartesian}
X_0 &=&  \pm \frac{1}{2} \eta \left(e^\psi -\epsilon
e^{-\psi}\right) \nonumber\\
X_1 &=& \pm \frac{1}{2} \eta\left( e^\psi +
\epsilon e^{-\psi}\right)\cos\theta \nonumber\\
X_2 &=& \pm \frac{1}{2} \eta\left( e^\psi +
\epsilon e^{-\psi}\right)\sin\theta\, .
\eea
Note that
\be
X^2 \equiv  - X_0{}^2 + X_1{}^2 + X_2{}^2 = \epsilon \eta^2\, ,
 \ee
Since $\eta^2$ is positive, it follows that $X^2<0$ when $\epsilon=-1$, and $X^2>0$
when $\epsilon=1$. The $X^2<0$ region is the Milne region of
Minkowski space and cosmological solutions are trajectories in this
space. Generic trajectories reach $\eta =0$ at finite FLRW time,
corresponding to a cosmological singularity.
However, the hypersurface $\eta =0$ is just the Milne horizon, and the
singularity at the Milne horizon disappears in the cartesian coordinates
$X_\mu $. The trajectory can therefore be smoothly continued through
the Milne horizon {\it in cartesian coordinates} to the region in which $X^2>0$,
where  we need $\epsilon=1$. Thus, on passing through the Milne horizon, a
cosmological trajectory becomes an instanton (and vice-versa).

The target space and the scale factor solutions given previously can
now be combined into a single solution for $X_\mu $. For example, for
$\ell^2>0$, the solutions are
\be
X_\mu =\cases{ \pm \eta \left[s_\mu \sinh(\ell \lambda )
+c_\mu \cosh(\ell \lambda )\right] \,, \quad & $\epsilon=1$ \,, \cr
 \pm \eta \left[s_\mu \cosh(\ell \lambda )
+c_\mu \sinh(\ell \lambda )\right] \,, \quad & $\epsilon=-1$ \,, }
\ee
where
\bea
s_0 &=& \sqrt{1+a^2}\cosh(\ell \lambda_0 )\ ,\ \ \ \ \ a\equiv
{q_- / \ell}\ ,
\nonumber\\
c_0 &=& -\sqrt{1+a^2}\sinh(\ell \lambda_0 )\ ,
\nonumber\\
c_1 &=& \cosh(\ell  \lambda_0)\cos(\theta_0)+a\sinh(\ell
  \lambda_0)\sin(\theta_0)\ ,\
\nonumber\\
s_1 &=& -\sinh(\ell
  \lambda_0)\cos(\theta_0)-a\cosh(\ell
  \lambda_0)\sin(\theta_0)\ ,
\nonumber\\
c_2 &=& -a\sinh(\ell
  \lambda_0)\cos(\theta_0)+\cosh(\ell
  \lambda_0)\sin(\theta_0)\ ,\
\nonumber\\
s_2 &=& a\cosh(\ell
  \lambda_0)\cos(\theta_0)-\sinh(\ell
  \lambda_0)\sin(\theta_0)\ .
\eea
Note that
$(c_\mu\pm s_\mu)$ is null.

\section{Instanton-cosmology transitions}
\setcounter{equation}{0}

The Milne gauge Lagrangian $L_\tau$ in cartesian coordinates is
\be\label{lefftau}
L_\tau = {1\over2}\left[ (dX/d\tau)^2  +
k\left(\epsilon X^2\right)^{\gamma-1}\right]\, .
\ee
The constraint is now
\be
(dX/d\tau)^2 = k \left(\epsilon X^2\right)^{\gamma-1}\, .
\ee
We thus have a problem analogous to that
of a particle of zero energy in a central potential,
with conserved $SL(2;\bR)$ ``angular momentum''
\be
\ell^\mu = \varepsilon^{\mu\nu\rho}X_\nu  (dX_\rho/d\tau)\, \ .
\ee
In contrast to the usual central potential problem, $X^2$ can be zero (or negative)
for non-zero 3-vector $X$, and $\ell^2$ may also be positive, negative or zero.
Nevertheless, the problem is still exactly soluble; the solutions
are the solutions $X_\mu(\lambda)$ given earlier with $\lambda$
expressed as a function of $\tau$. However, if $\eta\to 0$ as
$|\lambda|\to \infty$ it is generally necessary to piece together several of the
previously given solutions to obtain a full trajectory in the 3D
Minkowski space. In this section we shall show how this can be done.

The formulation of the problem in terms of cartesian field variables $X_\mu$
makes it obvious how trajectories can be smoothly continued through the
Milne horizon in certain special cases. For example, if $\gamma$ is an odd
integer then the Lagrangian (\ref{lefftau}) is independent of $\epsilon$. It then
follows from standard theorems about the existence and uniqueness
of solutions of ordinary differential equations that any trajectory that crosses
$\eta=0$  will connect smoothly to a solution on the
other side of the Milne horizon with the same value of $k$; a simple example is
$\gamma=1$, for which the trajectories are straight lines. We discuss this case in
detail below. If $\gamma$ is an even integer then the Lagrangian (\ref{lefftau})
depends on both $\epsilon$ and $k$ but only through the combination $\epsilon k$.
It follows that any trajectory that crosses $\eta=0$ must smoothly
join to a solution on the other side of the Milne horizon with a flipped sign of $k$.
A simple example is $\gamma=2$, which we also discuss in detail below.

It may not be obvious that a flip of the sign of $k$ is consistent with continuity in
the full field space, prior to imposing the ansatz (\ref{ansatz}), because one might expect
the $k=1$ and $k=-1$ metrics to belong to disjoint subspaces
in the space of all metrics. However, as previously observed,
the $k=\pm 1$ trajectories all approach a $k=0$ trajectory near
$X^2=0$, so the actual radius of curvature goes to infinity at a
cosmological singularity whatever the value of $k$. In other words, the subspace
of $k=1$ FLRW metrics is joined to  the subspace of $k=-1$ FLRW metrics
precisely at the points in the full space of fields at which we flip the sign of $k$, so
there is no discontinuity caused by this sign change.

Another  simple case  which we analyse in detail below is $k=0$, in which case
the results are obviously  $\gamma$-independent\footnote{
There is one other circumstance in which the physics
is independent of $\gamma$: it is obvious from (\ref{laginitial}) that solutions
with $\chi \equiv 0$ cannot depend on $b$ and hence that
solutions with $\theta\equiv 0$ cannot depend on $\gamma$. These are the
solutions with $X_2=0$.
It is not immediately obvious from the Lagrangian (\ref{lefftau}) why this should be the
case, but this can be seen from an application of the Jacobi principle
(see, e.g., \cite{Russo:2004am}), which states that zero-energy solutions of the equations of motion
of (\ref{lefftau}) are geodesics in the metric
$(\epsilon X^2)^{\gamma-1} dX\cdot dX$,
which is a conformal rescaling of the 3D Minkowski metric.
On the 2D subspace with $X_2=0$,
we can introduce lightcone coordinates $X_\pm $ and write the metric as
$(\epsilon X_-X_+)^{\gamma-1}dX_-dX_+$. Setting $ U=(\epsilon
X_-)^\gamma,\
V=\epsilon X_+{}^\gamma$, the metric becomes a constant times
$dUdV$, for any $\gamma$.}.

\subsection{The $k=0$ case: flat cosmologies }

For $k=0$, the equation of motion for $X^\mu$ is trivially solved by
\be\label{kazero}
X_\mu = a_\mu + p_\mu \tau\, ,
\ee
for constant 3-vectors $a$ and $p$. The constraint
implies that $p^2 =0$; in other words, the trajectories in Minkowski
space are null. In this case
\be
\ell^2= (a\cdot p)^2 \ ,
\ee
and it follows that $X^2 = a^2 \pm 2\ell\tau$. If $\ell \neq 0$ we are free to shift
$\tau$ to put this into the form
\be
X^2 =\pm 2\ell\tau\ .
\ee
Integrating (\ref{lamtau}) then yields
\be
\tau \propto  e^{\pm 2\ell\lambda}\, ,
\ee
and therefore
\be \label{eta-k=0}
\eta^2 \equiv \epsilon X^2 \propto e^{\pm 2\ell\lambda}\, .
\ee
Note that
\be
t \propto \pm \tau\, ,
\ee
whereas $\tau(\lambda)>0$. It follows, given
a choice of  sign,  that  the solution (\ref{one}) covers
only the part of the trajectory for which $t$ is either
positive or negative, but not both. Any null geodesic
must cross the horizon once, at $t=0$, so if we associate
$t>0$ with a big-bang cosmology
then we must associate $t<0$ with a pre-big-bang instanton.
These solutions give rise to the the upper and lower diagonal lines in
Fig.~\ref{fig:lines}; the two possibilities correspond to the
sign choice in (\ref{eta-k=0}). The transition from instanton
to cosmology, or {\it vice versa}, occurs at the hypersurface
$X^2 = 0$, which becomes the hyperbola
$X_0{}^2 - X_1{}^2 = X_2{}^2$ when projected onto the
$(X_0,X_1)$-plane.

The remaining possibility for $k=0$ is to have $a \cdot p=0$, i.e.~$\ell^2 = 0$.
In this case $X^2 = a^2$ and the geodesic will never reach the Milne horizon. This corresponds to the
middle diagonal line in Fig.~\ref{fig:lines}.

Thus for $k=0$ we have instanton-cosmology transitions with $\ell^2 > 0$ for any value
of $\gamma $. In particular, this is true for IIB supergravity (for
which $\gamma= 2/3$).

\subsection{The $\gamma=1$ case: geodesics}

The lagrangian $L_\tau$ is especially simple for
$\gamma=1$; in this case the equation of
motion for $X_\mu$ is solved by
\be\label{gamzero}
X_\mu = a_\mu + p_\mu \tau\, ,
\ee
for constant 3-vectors $a$ and $p$, and the constraint
implies that $p^2 =k$. Note that
\be\label{lvsa}
\ell^2 = (a\cdot p)^2 - ka^2\, .
\ee
As long as $k\ne0$ we can shift $\tau$, if necessary,
to arrange for $p\cdot a =0$, in which case $a^2= -k\ell^2$ and
\be
X^2 = k\left(\tau^2 -\ell^2\right) \,.
\ee
Note, however, that  $p\cdot a =0$ implies that  $\ell$ is non-spacelike
if $p$ is timelike; i.e., $\ell^2\ge0$ if $k=-1$.

\begin{figure}[ht]
\centerline{\epsfig{file=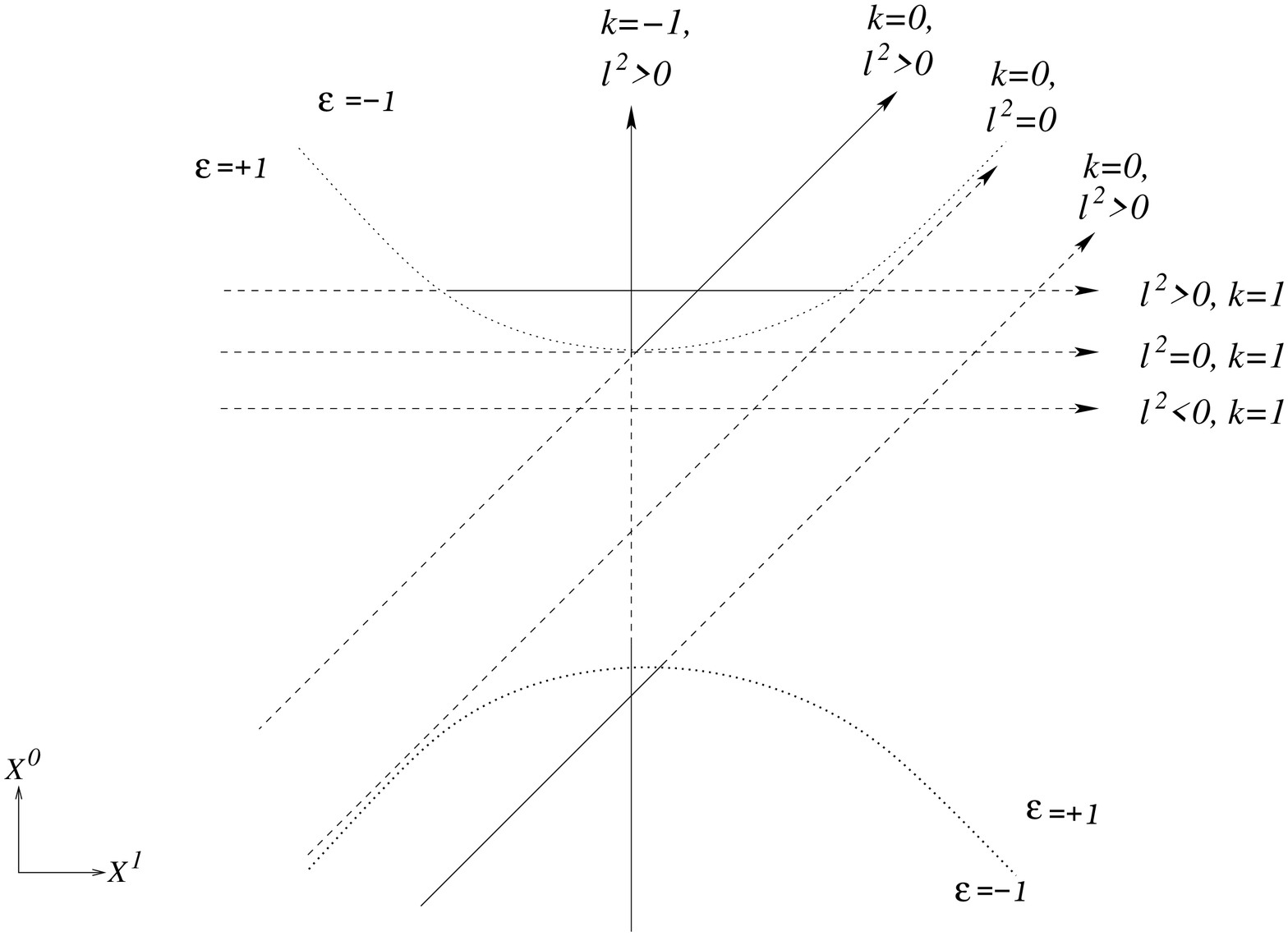,width=.9\textwidth}}
 \caption{\it Instantons and cosmologies as
geodesics in a projection of Milne space-time.
The solid (dashed) lines are cosmologies (instantons),
which are separated by the dotted hyperbola with
$X_0{}^2 - X_1{}^2  = X_2{}^2$.}
 \label{fig:lines}
\end{figure}

The $\ell^2>0$ cases are especially  interesting.
Consider first the $k=-1$ subcase, for which
$X$ is timelike for $|\tau|>\ell$ but spacelike
for $|\tau|<\ell$. In other words, a single
straight-line solution in Minkowski space can be
viewed as a cosmology ($\epsilon=-1$) for
$|\tau|>\ell$ but as an instanton ($\epsilon=1$)
for $|\tau| < \ell$. As argued
in \cite{Russo:2004am} (in the context of another
model with similar features) this
corresponds to a big-crunch/big-bang transition through
a classically forbidden region behind the Milne horizon.
For $k=1$ the roles of cosmology
and instanton are reversed, and the cosmological region
(in which $X^2<0$) corresponds to a
universe expanding from a big-bang (where the trajectory
first crosses the Milne horizon)
to a big-crunch (where it again crosses the Milne horizon).
For each  of these two
subcases, (\ref{lamtau}) reduces to
\be
d\lambda = {k\epsilon\, d\tau\over \tau^2-\ell^2} \,,
\qquad (\ell^2>0),
\ee
and we must solve this for $\tau(\lambda)$ in
two cases:
\begin{itemize}
\item $k\epsilon=-1$, $|\tau|<\ell$. In this case
\be
\tau = \ell \tanh \left(\ell\lambda\right)\, ,
\ee
and hence
\be
\eta^2 = X^2 = {\ell^2\over \cosh^2\left(\ell \lambda\right)}\, .
\ee
This is the $\gamma=1$ case of solution (\ref{two}).

\item $k\epsilon=1$. $|\tau|>\ell$. In this case
\be
\tau = -\ell\coth\left(\ell\lambda\right)\, ,
\ee
and hence
\be
\eta^2 =-X^2 = {\ell^2 \over \sinh^2\left(\ell\lambda\right)}\, .
\ee
This is the $\gamma=1$ case of solution (\ref{three}).
\end{itemize}
Thus, in these cases the full trajectory connects a
collapsing FLRW universe to an expanding FLRW universe
via an instanton solution; in fact, to one of the
super-extremal  D-instantons  of
\cite{Bergshoeff:2004fq}. As noted above, a change
of sign of $k$ reverses the roles of cosmology and
instanton, so which of the two ($k=\pm1$)  super-extremal
D-instantons is relevant depends on the sign of $k$.
Both these $\ell^2 > 0$ possibilities are illustrated in
Fig.~\ref{fig:lines}, where the time-like geodesics with
$k=-1$ are vertical lines while the space-like geodesics
with $k=+1$ are horizontal.

A similar calculation for $\ell^2<0$, for which we must set
$\epsilon=k=1$ for a non-trivial solution, yields \be \tau(\lambda)
= - \sqrt{-\ell^2} \, \cot\left(\sqrt{-\ell^2}\, \lambda\right) \,, \ee
and thence \be \eta^2 =  {-\ell^2\over \sin^2\left( \sqrt{-\ell^2}\,
\lambda\right)}\, . \ee This is the sub-extremal D-instanton
solution (\ref{timelike}), which is {\it periodic} in imaginary
time.

Finally, we consider $\ell^2=0$ (which includes $\ell_\mu=0$, and
hence $a_\mu=0$, as a subcase). In this case $X^2=k\tau^2$ and hence
$\tau = -(\epsilon k)/ \lambda$. This yields $\eta^2 \equiv \epsilon
X^2 = (\epsilon k)/\lambda^2$, which implies both  $\epsilon k=1$
and $\eta^2 = 1/\lambda^2$. This is the $\gamma=1$ case of the
solution (\ref{ellzero}). However, this solution must be interpreted
{\it either} as a cosmology (when $k=-1$) {\it or} as an instanton,
in fact the extremal D-instanton (when $k=1$) {\it but not both}.
For the interpretation as a $k=-1$ cosmology we noted previously
that this case yields a $d$-dimensional Milne universe\footnote{This
is a {\it spacetime} Milne universe, and is not to be confused with
the Milne `universe' in which cosmological trajectories evolve.} for
which the apparent cosmological singularity is actually just a
coordinate singularity that  can be resolved without the need for
scalar fields.Thus, we should not expect (and have not found) any
`instanton phase' of the cosmological trajectory in this case. The
cosmological and instanton interpretations are, in this case, {\it
separate} solutions, not linked by a spacetime cosmological
singularity. In Fig.~\ref{fig:lines} we have illustrated the $k =
1$ instanton with $\ell^2 = 0$. The corresponding cosmological
solution with $k=-1$ is not indicated but consists of the $\ell
\rightarrow 0$ limit of the $\ell^2 > 0$ case. This limit implies
that $X_2 \rightarrow 0$: the two hyperbola join to form a cross and
the 'instanton phase' of the solution disappears.

We have now seen how {\it all} the solutions found in
section \ref{sec.exact} correspond, for $\gamma=1$, to
some straight-line trajectory in the 3-dimensional
Minkowski spacetime with coordinates $X_\mu$.
Note, in particular, that a single straight line trajectory
in Minkowski `superspace' corresponds to cosmological
and instanton solutions with the {\it same} value of $k$.

\subsection{The $\gamma=2$ case: cyclic universes}
There is one other value of $\gamma$ for which the
equations we have to solve are {\it linear}
in Minkowski coordinates, namely $\gamma=2$. We see from
(\ref{lefftau}) that the effective Lagrangian in this case is
\be\label{lefftau2}
L_\tau = {1\over2}\left[ \dot X^2  +
(k\epsilon) X^2 \right]\, .
\ee
The equations of motion are
\be
\ddot X_\mu  = (k\epsilon) X_\mu \,, \qquad (\mu=0,1,2).
\ee
We must choose solutions that satisfy the constraint
\be
\dot X^2 = (k\epsilon)\, X^2\, ,
\ee
which can be interpreted as a `zero-energy' condition. We have already discussed
the $k=0$ case, so we may assume that $k\ne 0$. We now consider in turn the two
possible values of  $k\epsilon$:
\begin{itemize}

\item $k\epsilon=1$. In this case, the equations are solved by
\be
X_\mu = A_\mu e^\tau + B_\mu e^{-\tau} \, ,
\ee
for real 3-vectors $A,B$,  and the constraint implies that
\be
A\cdot B=0\, .
\ee
This implies that
\be
X^2 = A^2 e^{2\tau} + B^2 e^{-2\tau}\, ,
\ee
and that
\be
\ell^2 = -4 A^2B^2\, .
\ee
If both $A$ and $B$ are null then we have a solution with $\ell^2=0$ and
$X^2=0$. This does {\it not} correspond to one of the solutions found in
section \ref{sec.exact} because the coordinates used there do not cover
the hypersurface $X^2=0$, but this `extra' solution is of no physical interest. For
the other  $\ell^2=0$ cases, we may assume without loss of generality
that  only $A$ is null, so that
\be
X^2 = B^2e^{-2\tau}\, ,
\ee
where $B$ is non-null.  If $A$ is non-zero  then $B$ must be
spacelike in order to be orthogonal to $A$, in which case $X^2>0$. This
case is relevant only for $\epsilon=1$, in which case $k=1$. These are extremal
D-instantons for $\gamma=2$. If $A=0$ then $B$ may be either spacelike or
timelike. If $B$ is spacelike then we must choose $\epsilon=1$ and we then
have a further special case of the extremal D-instanton. If $B$ is timelike then
we must choose $\epsilon=-1$ and we then have the unique ($k=-1$) $\ell^2=0$
FLRW cosmology, for which $\eta \propto e^{-\tau}$. This is non zero for all finite
$\tau$, so it might appear that this is a cosmology without a singularity. However,
as we shall see in the following section, the $\gamma=2$ case is a very special
one for  $\ell^2=0$ because $\tau$  becomes infinite for finite FLRW time $t$.

This leaves those cases for which neither $A$ nor $B$ is null. If both are timelike or
spacelike then  $\ell^2<0$ and  $X^2$ is never zero. If one is timelike and the other spacelike
then $\ell^2>0$ and $X^2$ passes through zero; this is case (\ref{two}) of
subsection \ref{subsec.scale}.

\item $k\epsilon=-1$. In this case,
\be
X_\mu = C_\mu e^{i\tau} + \bar C_\mu e^{-i\tau}\, ,
\ee
where $C_\mu$ is a complex 3-vector (with complex
conjugate $\bar C_\mu$) subject to the constraint
\be
C\cdot \bar C=0\, .
\label{cco}
\ee
This implies that
\be
X^2 = C^2 e^{2i\tau} + \bar C^2 e^{-2i\tau} \,,
\ee
and that
\be
\ell^2 = |2C^2|^2 \ge 0\, .
\ee
For $\ell^2=0$ (which occurs when $C$ is null) we again have the
$X^2=0$ case. Otherwise, $\ell^2>0$ and $X^2$ passes through zero whenever
\be
\tau = \tau_0 + n\pi/2\, ,\qquad (n\in \bZ),
\ee
where $\tau_0$ is any solution of $e^{4i\tau} = - \bar C^2/C^2$.

\end{itemize}

As anticipated, all non-trivial cosmological trajectories with
$\ell^2>0$ have an instanton phase, behind the Milne horizon, to which
they are smoothly connected through a cosmological singularity. These trajectories
connect $k=\pm1$ cosmologies with $k=\mp1$ instantons, and in the
$k\epsilon=-1$ case the trajectories are cyclic universes. We
conclude this subsection with an explicit  example.

First, note that the constraint (\ref{cco}) is solved,
without loss of generality, by
\be
C_\mu = A \left( 1, e^{i\beta_1}\cos \xi , e^{i\beta_2}\sin\xi\right)\, ,
\ee
where $A$ is a real constant determined by $|2C^2|=\ell$, and $\beta_1,\beta_2,\xi$ are three real constant angles. This yields
\be
X_\mu = 2A\left(\cos\tau, \cos \left(\tau+\beta_1\right)\cos\xi ,
\cos\left(\tau+\beta_2\right) \sin\xi\right)\ .
\ee
Consider the particular case $\beta_1=-{\pi\over 2}$, $\beta_2=0$, for which
\be
X_\mu = {\sqrt{\ell}\over \cos\xi} \left(\cos\tau, \sin\tau \cos\xi ,
\cos\tau\sin\xi \right)\ .
\ee
In this case  $X^2 =- \ell\cos 2\tau$, independent of $\xi$. Recalling that $\epsilon X^2\ge 0$ for all $\tau$, we see that
\be
\epsilon X^2 = \ell\left|\cos 2\tau\right|\, .
\ee
Using this in $d\tau = \eta^2d\lambda$, we may
integrate to deduce that $\left|\cos 2\tau\right| = 1/\cosh 2\ell\lambda$, and hence that
\be
\epsilon X^2 = {\ell\over \cosh 2\ell\lambda}\, .
\ee
This is precisely the $\gamma=2$ case of (\ref{two}), where for $\epsilon=-1$
we viewed it as a big-bang to big-crunch $k=1$ cosmology. For this interpretation
we should choose $\tau \in (-\pi/4,\pi/4)$ but we  have now
discovered that  we can follow this solution through the big-crunch at $\tau= \pi/4$
to an instanton solution, given by the same formula but with
$\tau \in (\pi/4, 3 \pi/4)$. This then re-emerges as  another big-bang cosmology for
$\tau\in (3\pi/4, 5\pi/4)$.  If the first one evolved in
the future-Milne region of the Minkowski `superspace' then the second evolves in
past-Milne region. After a further recollapse to an instanton phase for $\tau\in (5\pi/4, 7\pi/4)$,
the cycle is completed as this gives way to a new big-bang universe. There are thus two
big-bang and two big-crunch singularities in every cycle, but the transition
through them is smooth  in the augmented target space (see Fig.~3).

\subsection{Generic $\gamma$}

As already observed, the Lagrangian (\ref{lefftau})  tells us that for general
$\gamma$ we have a problem
analogous to that of a particle in a central potential. Using (\ref{cartesian}) to
return to `polar' coordinates, we
find that the Lagrangian governing the radial motion is
\be
L_{rad} = {\epsilon\over2}\left[  \left({d\eta\over d\tau}\right)^2 +  {\ell^2\over\eta^2} +
(k\epsilon) \eta^{2(\gamma-1)} \right] \,.
\ee
As expected, we have a `centrifugal'  term for non-zero $\ell^2$.
Recall that the constraint implies that we must retain only the `zero-energy' solutions
of the equation of motion, for which
\be\label{contau}
\left({d\eta\over d\tau}\right)^2 = {\ell^2\over\eta^2} + (k\epsilon) \eta^{2(\gamma-1)}\, .
\ee
This is equivalent to (\ref{con1}) but with $\tau$ as the independent variable. The solutions are
therefore the same as those given in \ref{subsec.scale} but with $\eta$ expressed as a
function of $\tau$ rather than $\lambda$.

For $\ell^2<0$ we have a centrifugal barrier that prevents $\eta$ from
passing through zero. For $\ell^2>0$ the centrifugal barrier becomes
a `centrifugal well' that dominates near $\eta=0$.
For $k\epsilon=-1$ the potential  becomes positive for sufficiently
large $\eta$; as positive values of the potential  are not accessible for
zero energy, the `particle' is confined to finite $\eta$, and must fall to
$\eta=0$ because the potential has no stationary points (under the
assumed conditions). For $\ell^2>0$ and $k\epsilon=1$ the potential is
always negative; it has a stationary point if $\gamma>1$ but the energy
is fixed at zero so all trajectories must start at infinite $\eta$ and then fall
to $\eta=0$. The net conclusion is that $\eta$ reaches zero
on {\it all} $\ell^2>0$ trajectories, in agreement with our cartesian coordinate
analyses for $\gamma=1$ and $\gamma=2$. Moreover, from the
explicit $\ell^2>0$ solutions (\ref{one})-(\ref{three}), and the relation
 $\eta^2d\lambda=d\tau$, one may show that
\be
\eta^2 \propto |\tau| \propto |t|^{1/(\gamma\alpha^2)} \,, \qquad
{\rm as}\ \eta\rightarrow 0\, , \qquad (\ell^2>0) \,,
\ee
where $t$ is the FLRW time, and here we have chosen the time origin
such that $t\rightarrow 0$ as $\tau\rightarrow 0$. This shows that
there is a cosmological singularity at $t=0$, corresponding to the
time $\tau=0$ at which the cosmological trajectory reaches $\eta=0$.

The $\ell^2=0$ case is special and needs a separate  discussion. In this
case we may assume that $k\epsilon=1$ because otherwise there is no
solution other than the trivial $k=0$ solution for which $\eta$ is
constant. Given $k\epsilon=1$, there is a negative effective
potential and hence no obvious barrier to prevent $\eta$ passing
through zero. However, the $\ell^2=0$ solution is
 \be
  \eta =  \cases{
   [ (\gamma-2)\tau ]^{1/(2-\gamma)} \,, \quad \gamma \neq 2 \,, \cr
   e^{-\tau} \,, \qquad \qquad \qquad \; \, \gamma = 2 \,.}
 \ee
This shows that $\eta$ reaches zero {\it at
finite $\tau$} if $\gamma<2$ and {\it at infinite $\tau$} if $\gamma \geq 2$ (in agreement
with our earlier analysis of the $\gamma=1$ and $\gamma=2$ cases).
This does not necessarily mean that $\eta$ will not pass through
zero if $\gamma \geq 2$ because it may happen that $\tau$ becomes
infinite for finite FLRW time $t$. To see whether this happens we
need to consider the relation between $t$ and $\tau$, which is
 \be
dt = \eta^{2(\gamma\alpha^2 -1)} d\tau\, .
 \ee
This yields for all values of $\gamma$
 \be
\eta = \left( \frac{\gamma t}{2-d} \right)^{(d-2)/\gamma} \,, \qquad (\ell^2=0) \,.
 \ee
One sees from this result that for all values of $\gamma$ there is a cosmological
singularity at $t=0$, i.e.~{\it at finite FLRW time}.
However, as already noticed for $\gamma=1 $, there is no transition
from a cosmology to an instanton, or vice-versa, when $\ell^2=0$.

For $\gamma=1 $ and $\gamma=2$, we have seen that in Minkowski
field variables there is actually a smooth connection at $\eta=0$ onto
an instanton solution in the region behind the Milne horizon. We now
want to determine whether a similar smooth transition is possible for other
values of $\gamma$. As we have just seen, this issue arises only for
$\ell^2>0$, so we now restrict our discussion to that case. From the general
solutions found previously we see that a trajectory can reach $X^2=0$
only as $|\lambda|=\infty$. Let us concentrate on the case in which
$\eta\to 0$ as $\lambda\to \infty$; in this case\footnote{Recall that the
parameter $\lambda$ in the $X^2>0$ region is independent of the
parameter $\lambda$ in either of the $X^2<0$ regions.}
\be
\lim_{\lambda\to \infty} X_\mu(\lambda) = c_\mu + s_\mu \ ,
\ee
independently of $k$. It  follows that  the $\epsilon=-1$ solution
for $X_\mu$ can be matched continuously at  $\lambda=\infty $
onto the  $\epsilon=1$ solution at $\lambda=\infty$, {\it without
changing parameters}; i.e., with the same $\lambda_0$, $q_-$ and $\theta_0$
for the instanton and cosmological solution.

The next question is whether the transition is continuous for
the first derivatives. Using the formulas given above, we find the
following results\footnote{The $\gamma $-dependence arises because the
  ($\gamma$-independent) leading terms at large $\lambda $ of
  ${dX_\mu\over d\lambda }$ cancel.}:
\be
\lim_{\lambda\to \infty} {d X_i\over d X_0}=\cases{
(c_i + s_i)/(c_0 + s_0) & $\gamma < 1 $ \cr
c_i / c_0  & $\gamma = 1,\ k=1 $ \cr
s_i / s_0  & $\gamma = 1,\ k=-1 $ \cr
(c_i- s_i)/( c_0- s_0) & $\gamma > 1 $ } \qquad (i=1,2).
\ee
Thus, the first derivatives match in all cases with the same choice of
$\theta_0$, $q_-$ and $\lambda_0$ parameters on both sides.

Note that for $\gamma=1$ continuity of the first derivatives implies that one must
patch together solutions with the same value of $k$, leading to the straight lines of
section 3.1. For all other values of $\gamma$ we could choose to patch together solutions
with the same or opposite $k$, but  continuity of second derivatives
imposes further restrictions. From the general solutions given previously,
one can show that
\be
\lim_{\lambda\to \infty} {d^2 X_i\over d X_0^2}=\cases{
0 & $\gamma <{1/ 2}$ \cr
\epsilon\times {\rm const.}  & $\gamma ={1/ 2}$ \cr
\epsilon\times \infty   & ${1\over 2}<\gamma < 1$ \cr
0 & $\gamma =1 $ \cr
\epsilon k\times \infty   & $1<\gamma < 2$ \cr
\epsilon k\times{\rm const.}& $\gamma =2 $ \cr
 0& $\gamma >2 $ \cr
} \qquad (i=1,2).
\ee
This shows, for instance, that when $1<\gamma\leq 2 $ the second derivatives are
continuous if $k \epsilon $ is the same on both sides, meaning that on a smooth
trajectory $k$ must flip sign across the Milne horizon; otherwise there is a discontinuity  in
the second derivatives. For $\gamma>2$ the second derivative is continuous
irrespective of whether $k$ flips sign; moreover, as already observed, there is an
analytic extension for integer $\gamma$. For $\gamma<1/2$ there is also no discontinuity
in second derivatives whether or not $k$ flips sign. However, when ${1\over 2}\leq \gamma < 1$, there is a discontinuity in second derivatives  irrespective of whether $k$ flips or remains the same, which means that there is no smooth transition through the Milne horizon for $\gamma$ in this range.

\begin{figure}[two]
\centerline{\epsfig{file=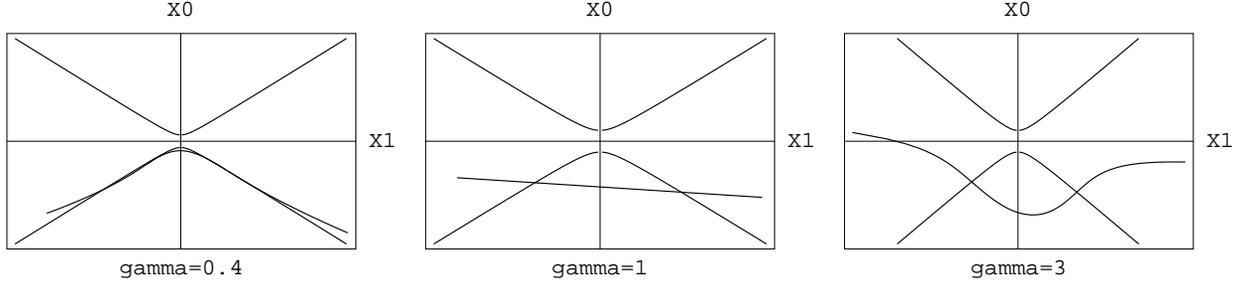,width=1.2\textwidth}}
 \caption{\it Instanton extensions of closed big-bang to big-crunch universes
 for $\gamma=0.4$, $\gamma=1$ and $\gamma=3$. The $\gamma=1$ curve is a straight line
 and the $\gamma=3$ curve is analytic.}
 \label{fig:geodev}
\end{figure}

To summarize: there are transitions between instanton and cosmological `phases'
for $k\ne0$ that are continuous up to and including second derivatives provided that
$\gamma<1/2$ or $\gamma\ge1$. For example, a $k=1$ universe, which would normally
be thought to start with a big-bang and end with a big-crunch,
may actually be part of a larger instanton-cosmology with pre-big-bang and post-big-crunch
instanton phases, as illustrated  in Fig.~\ref{fig:geodev} for $\gamma=0.4$, $\gamma=1$ and
$\gamma=3$. The $\gamma=0.4$ curve is probably not analytic.  In general, we would expect
continuity in all derivatives to impose further restrictions, and it may be that an analytic continuation is possible only for integer $\gamma$.

\begin{figure}[three]
\centerline{\epsfig{file=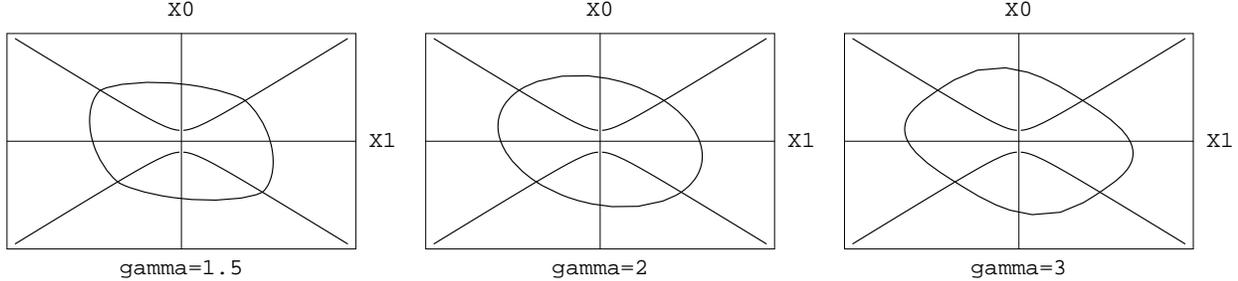,width=1.2\textwidth}}
 \caption{\it Cyclic universes with $k \epsilon = -1$ for $\gamma>1$. The generic planar curve for $\gamma=2$ is
an ellipse.}
 \label{fig:cyclic}
\end{figure}

As we have seen for the $\gamma=2$ case, these
instanton-cosmology transitions  can connect to form cyclic universes, and this remains true for all even $\gamma$; in fact, it is true for all $\gamma>1$ if one requires only continuity up to
and including second derivatives. This is  illustrated for {\it planar} trajectories in Fig.~\ref{fig:cyclic}.

\subsection{Big-bounce universes for IIB supergravity}

There is an important feature that distinguishes the trajectories in
$\gamma<1$ models from those in  $\gamma>1$ models. In both cases,
a trajectory that approaches the Milne
horizon is null at the point of crossing. For $\gamma<1$ this null
curve approaches a null geodesic generator of the cone $X^2=0$, since
$dX^\mu \propto X^\mu$. In contrast, for $\gamma>1$, $dX^\mu $ is not
proportional to $X^\mu $.

This means that there is an additional possibility for $\gamma <1 $:
in this case a cosmological ($\epsilon=-1$) trajectory with $k=\pm 1$ may be smoothly
joined to another {\it cosmological} ($\epsilon=-1$) solution $k=\mp
1$, where by `smooth' we mean continuity up to and including second
derivatives.

\begin{figure}[three]
\centerline{\epsfig{file=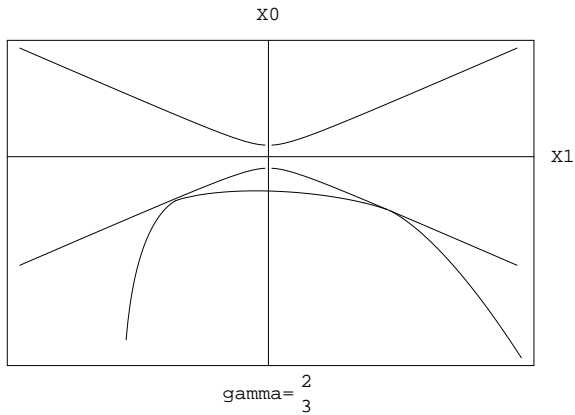,width=.5\textwidth}}
 \caption{\it A big-bounce universe for $\gamma=2/3$, in which an open universe
 that is collapsing to a big-crunch bounces off the Milne horizon in the augmented
 target space to emerge as a closed big-bang universe. The latter necessarily collapses
 to a big-crunch singularity, but this is again another bounce
 off the Milne horizon from which there emerges an open, but expanding, universe.}
 \label{fig:bounce}
\end{figure}

For example, a collapsing closed universe can bounce off
its big-crunch singularity to begin another phase
as an expanding open universe. This possibility is illustrated
in Fig.~\ref{fig:bounce} for $\gamma=2/3$.

This possibility is relevant to IIB supergravity since $\gamma= 2/3$ for the
uncompactified $d=10$ theory. By considering a Kaluza-Klein ansatz of the form
 \be
  ds^2= ds^2_{d} + dy_n^2 \,, \quad \phi = \phi_d \,, \quad \chi = \chi_d \,, \quad (n=d+1,...,10) \,,
 \ee
we get an effective action in $d<10$ dimensions with\footnote{
 By embedding the lower-dimensional scalars differently one can obtain other values
 of $\gamma$, however. For example, $N=8$ supergravity in four dimensions can be truncated to gravity coupled to a vector and a scalar with dilaton couplings $a=0,1/\sqrt{3},1,\sqrt{3}$ \cite{Hull:1994ys}. Upon reduction to $d=3$ these give rise to our model (\ref{laginitial}) with $\gamma=1, \sqrt{3}/2,1/\sqrt{2},1/2$. Note that these are all in the range
$1/2 \leq \gamma \leq 1$.}
 \be
  \gamma={1\over 2} \sqrt{\frac{2(d-2)}{(d-1)}}\ .
 \ee
Assuming that $d\ge 3$, this implies that $\gamma\in [1/2,2/3]$.
This is precisely in the range ${1\over 2}\leq \gamma<1$ for which a
smooth transition to an instanton is not possible (for $k \neq 0$). However, a
smooth bounce {\it is} possible, as we have just seen.

\section{Discussion}
\setcounter{equation}{0}

In this paper we have expanded on the observation
in \cite{Russo:2004am} that cosmological singularities
may be resolved in models with scalar fields
parametrizing a hyperbolic target
space via an interpretation as coordinate singularities
of a Milne `super-metric' on the `augmented target space' of
scalar fields {\it and} metric scale factor. Specifically, the segments of cosmological
trajectories that lie behind the Milne horizon (for which
the scale factor of the Lorentzian-signature spacetime
is imaginary) have been shown to correspond, in a
particular class of models  with $H_2$ target space, to the
`super-extremal' D-instanton solutions of \cite{Bergshoeff:2004fq}.

It is interesting that the original, extremal, D-instanton of
\cite{Gibbons:1995vg}, generalized to  $d$ spacetime dimensions,
does not connect to a cosmological solution in the way that the
super-extremal D-instanton does, and the same is true of the
sub-extremal instantons of  \cite{Bergshoeff:2004fq}. The reason for
this is that the extremal and sub-extremal  instantons are
non-singular, so there is no way to join them on to any cosmological
solution in a continuous way. Recalling the debate over the
significance of singularities in the  context of the Hawking-Turok
cosmological instanton \cite{Hawking:1998bn}, it is amusing to note
that singularities of D-instantons are essential to the cosmological
interpretation that we have proposed for  them.

The models considered here generalize  the  metric-scalar sector of IIB
supergravity to arbitrary spacetime dimension $d$ and an arbitrary
$H_2$ radius, which is simply related to the parameter $\gamma$
arising in our analysis. The $\gamma=1$ case is particularly simple
as the cosmological-instanton trajectories are just straight lines.
In particular, the $k=-1$ universes of this model undergo a
big-crunch/big-bang transition of the type proposed in
\cite{Russo:2004am}, for which we have here identified
the intermediate `instanton phase' . Another special case is $\gamma=2$,
for which closed universes allow a smooth continuation to cyclic universes
corresponding to closed curves in the analytic extension of the
augmented target space. The idea that the universe
may be cyclic is an old one that has recently been revived in the braneworld
approach to cosmology \cite{Steinhardt:2001vw}; here we
have found an explicit model that realizes a cyclic universe.

For $\gamma<1/2$ and $\gamma \ge1$, corresponding to two disjoint ranges
of the $H_2$ radius, we have shown that there exists a continuation of $k \neq 0$ 
cosmological trajectories  through the Milne horizon with continuous first
and second derivatives, and the continuation is analytic for integer
$\gamma$. For the values $1/2 \leq \gamma < 1$, which are relevant for IIB supergravity and its
compactifications to $d<10$, we found that there is a smooth `big-bounce'
solution in which a collapsing cosmology is bounced off the big-crunch
to become an expanding big-bang universe. For $k=0$ there is a
smooth transition through the Milne horizon for any $\gamma$ but this
transition occurs just once, yielding either a pre-big-bang instanton
phase of a flat expanding universe or a post-big-crunch instanton phase
of a flat collapsing universe.

Our results are also of potential relevance to IIA string theory because in
the special case of zero axion, i.e. $\chi = 0$, the lagrangian (\ref{laginitial}) is
a truncation of massless  IIA supergravity, or its Euclidean counterpart.
Interestingly, there is an instanton/cosmology solution that survives this truncation,
which corresponds to the hypersurface $X_2 = 0$ in cartesian coordinates;
it is just the $q_-=0$ solution (\ref{special}) with the scale factor $\eta$ given by
eqs. (\ref{one}) - (\ref{three}). For $d=10$ and $\epsilon = k = 1$ this is the non-extremal
IIA D-instanton of \cite{Harvey:2000qu}; it has an M-theory origin
because the Kaluza-Klein ansatz
\be
ds^2_{11} = e^{2\alpha\phi/(d-1)}ds^2_{10} -\epsilon e^{-\phi/\alpha}dx_{10}^2
\ee
takes the $d=11$ Einstein-Hilbert Lagrangian to the zero axion truncation of
 (\ref{laginitial}) for $d=10$. One can thus show that  both `phases' of
 the IIA instanton/cosmology solution have a  common M-theory origin as a Schwarzschild black
 hole: the IIA instanton is obtained by reducing the black hole exterior spacetime over time
\cite{Harvey:2000qu,Bergshoeff:2004fq} while the  IIA cosmology  is obtained by reducing the interior
spacetime over a space direction \cite{Behrndt:1994ev}. An interesting open question
is whether there is an extension to the massive IIA theory, for which there is a scalar potential
for the dilaton.  We hope to show in a future paper how the results obtained here generalize to models
with a scalar potential that include as special cases both massive IIA supergravity and the model
with cosmological constant that was used in \cite{Russo:2004am}  to study the big-crunch
to big-bang transition for flat cosmologies.

\section*{Acknowledgements}

E.B. and A.C.  are supported
by the  European Commission FP6 program
MRTN-CT-2004-005104 in which E.B.
and A.C. are associated to Utrecht University.
The research of D.R. is funded by the PPARC grant PPA/G/O/2002/00475. He
would also like to thank the Centre
for Theoretical Physics of the University of Groningen for hospitality
during part of this work. The authors thank Ulf Gran, Michael Green and Sijbo Holtman
for stimulating discussions.

\setcounter{section}{0}
\bigskip



\end{document}